\RequirePackage{fix-cm}
\documentclass[smallextended]{svjour3}       
\smartqed  
\usepackage{graphicx}
\usepackage[colorlinks,citecolor=blue]{hyperref}
\usepackage{cite}
\usepackage{amsmath}
\usepackage{amsfonts}
\usepackage{amssymb}
\usepackage[caption=false,font=footnotesize]{subfig}
\journalname{Wireless Personal Communications}
\begin{document}
\title{Multi-objective Antenna Selection in a Full Duplex Base Station}

\author{Mohammad Lari \and Sina Asaeian}
\institute{M. Lari \at
	Electrical and Computer Engineering Faculty, Semnan University, Semnan, Iran \\
	\email{m\_lari@semnan.ac.ir}}

\maketitle
\date{Received: date / Accepted: date}

\begin{abstract}	
The use of full-duplex (FD) communication systems is a new way to increase spectral efficiency. For this reason, it has received serious attention in the new generation of wireless communication systems. The main challenge of FD systems is self-interference that needs to be reduced appropriately. In this paper, we have considered an FD multi-antenna base station in a cellular network and we have used the antenna selection technique to resolve the self-interference issue. We have also provided a new criterion to select an appropriate antenna. In this new criterion, the antenna selection is modeled as a multi-objective optimization problem. Here, the antennas which simultaneously minimizes interference channel gain and maximizes uplink (UL) and downlink (DL) channel gains are selected for transmission and reception. The base station has to perform well in both the UL and DL and reduce the self-interference simultaneously. Therefore, the multi-objective criterion has a better performance than the single-objective criterion. Although, in conventional antenna selection with the single-objective criterion, only the function of the UL and DL channels or the interference channel is considered. Finally, the simulations show that the new criterion has a higher throughput rate than the other conventional single-objective antenna selection techniques.
\end{abstract}

\section{Introduction}\label{sec:introduction}
In recent years, and especially with the growing use of smart equipment and phones, we see a superb increase in users’ traffic in telecommunication networks. Field research suggests that data traffic in the next decade will be about 1000 times the current level \cite{Ref01}. To this end, the new generation of communication systems has considered various and different plans to respond to this volume of requested traffic. Proposed solutions include miniaturizing and condensing network cells, using new and unutilized frequency bands as well as new waveforms to increase spectral efficiency and better utilization of available resources \cite{Ref01,Ref02}. One of the new waveform is the full-duplex (FD) transmission and reception. This trick is a new method in comparison to half-duplex (HD) transmission. Due to the increase in spectral efficiency in FD communications compared to HD, this method has considered in this paper.

In all telecommunication systems, there are two or more nodes that these nodes are usually exchanging data together one by one. A node sometimes acts as a transmitter and sends data to another node. The same node at another time plays the role of a receiver and receives data from the other node. Therefore, most telecommunication nodes have both transmitter and receiver. In current communication systems which work as HD, the transmission and reception of a node occur at different time slots or at different frequency bands. In the first case, the node sends data in a certain time slot and receives data in another time slot too. Also, in the second case, although the node sends and receives at the same time, but transmits and receives at a different frequency bandwidth and there is no interference in simultaneous transmission and reception. However, in FD systems, sending and receiving a node is done simultaneously in the same frequency bandwidth. In this way, when the receiver of a node is on and receives data, its transmitter also works at the same time and sends data. Therefore, the output power of the transmitter, which is usually strong, can easily leak into the receiver and create self-interference. The self-interference may even reach up to 100dB higher than the noise power at the receiver. For this reason, self-interference is exactly the cause that communication systems have not yet been implemented as FD \cite{Ref03,Ref04}. Currently, due to the researches done and the practical experiments made in the laboratory, it has been possible to reduce this interference to a certain extent within the receiver noise power. For this reason, utilizing the FD method has been proposed for new generations of communication systems \cite{Ref05}.

The main problem with the FD systems is the intense self-interference from the transmitter to the receiver of the same node. There are various methods to reduce this interference. Reducing interference is often done in the propagation domain, in the analog domain in the radio frequency (RF) section of the receiver or in the digital domain \cite{Ref04,Ref05}. Due to the high severity of this interference, practical systems should use the appropriate methods in all three areas of propagation, analog, and digital domains, simultaneously. So, in the end, the remaining interference reaches to (or less than) receiver noise floor \cite{Ref05}. Then, the FD system can send and receive simultaneously without any problems in a similar frequency band. Therefore, to communicate with another node, compared to an HD system, FD systems requires a shorter time slot and lower frequency bandwidth. This means (almost) doubling the spectral efficiency that is highly desirable for the new generation of communication systems.

Interference cancellation in the propagation domain is very important. Because the intensity of the interference is high and if this self-interference gets to the RF section of the receiver, causes saturation of RF circuits and completely interrupts receiver's functionality \cite{Ref06}. The methods for reducing interference in the propagation domain often include antenna selection \cite{Lari01, Lari02}, the use of electromagnetic absorbers between transmit and receive antennas \cite{Ref07} and the use of orthogonal polarizations \cite{Ref08,Ref09} for sending and receiving with the least effect of coupling. The propagation domain techniques are not able to remove the self-interference completely. Therefore, before converting the signal to digital values, another portion of the interference should be removed in the analog domain. Otherwise, due to the limited dynamic range of the analog to digital converter, some parts of the data may be lost. Due to the limited processing gain that exists in the analog section of the receiver, the elimination of interference in this section often involves estimating the interference and subtracting it from the received analog signal \cite{Ref10}. Then, the analog signal converts to digital values and the remaining self-interference in the digital domain diminishes greatly. Due to the high processing gain in the digital section, digital interference cancellation techniques are very diverse. Among them, we can point out the interference alignments \cite{Ref11} and the methods based on subspace \cite{Ref12,Ref13}. In most of these methods, the degree of freedom in multi-antenna systems are used to separate the interference space from the desired signal.

In addition to self-interference cancellation, the antenna selection can also simplify and reduce the volume and power consumption of the RF section in both transmitter and receiver \cite{Lari03}. Hence, the use of this method is very common in multi-antenna communication systems. This article also used this trick to reduce interference. Several articles have addressed this issue. One of the best in this field is \cite{Ref14} in which the FD system is considered as multi-antenna and by using an antenna selection or beam selection, the interference is reduced and, even in some situations, has been completely canceled. This method is also used in \cite{Ref15} with relay selection in amplify and forward (AF) relays to reduce interference. The authors of \cite{Ref15} also calculated the outage probability. Antenna selection in a multi-antenna FD communication system is also explored in \cite{Ref16} and \cite{Ref17} where two nodes are exchanging data simultaneously and within the same frequency band. In \cite{Ref16}, the self-interference is reduced by selecting one antenna in the transmitter and one antenna in the receiver of the first node and in the same way one antenna in the transmitter and one antenna in the receiver of the second node. Then the authors calculate the average sum rate and the average sum symbol error rate. In \cite{Ref17} which is more general than \cite{Ref16}, the number of selected antennas per node for sending and receiving is not necessarily equal to one and can be higher. In this case, authors with appropriate approximations have calculated the average sum rate of the bi-directional communication system. In addition, different criteria for selecting an appropriate antenna at a base station in a cellular communication network that works as an FD is studied in \cite{Ref18} and the outage probability was calculated in both downlink (DL) and uplink (UL) paths.

The FD base station, which is our subject of the article, has the ability to send data for a user terminal in the DL path and receive data from another user terminal in the UL path simultaneously in the same frequency bandwidth. Obviously, if the FD base station reduces its self-interference as much as possible and provides service to both DL and UL users at one time slot in the same frequency band, spectral efficiency will increase. In the cellular network, the use of FD base station has several advantages. One is that the processing unit at the base station is usually strong and powerful. Therefore, the complex interference cancellation technique can be implemented in the base station more easily. Another advantage is that when the base station is implemented as an FD, there is the possibility of increasing spectral efficiency without changing the terminal equipments. Even, the terminals do not even need to know about the functionality of the base station as FD. According to this description, here, the base station of a cellular network is considered to be FD, and the antenna selection has been used to reduce self-interference in it. In the following, the two antenna selection criteria which are studied in \cite{Ref18} will be discussed and the disadvantage of those two criteria will be described. Then, a new criterion for selecting the antenna at the base station will be presented. Since the goal of the base station is to provide the best service to users in both DL and UL paths with the least self-interference, so in general, we are facing a multi-objective optimization problem. Therefore, the proposed criterion for antenna selection has also been designed and solved as a multi-objective optimization. Finally, with simulations, it will be shown that the new criterion has a good performance, especially in comparison with the common antenna selection techniques.

In the following, first, the system model will be described in section \ref{sec:system_model} and the common criteria for antenna selection at the base station will be stated. Then, in section \ref{sec:multi_objective_criterion}, the multi-objective criteria for antenna selection are explained, and in section \ref{sec:simulation_results} the simulation results are presented. Finally, in section \ref{sec:conclusion}, the conclusion is given.

\section{System Model}\label{sec:system_model}
\begin{figure}
	\includegraphics[width=0.7\linewidth]{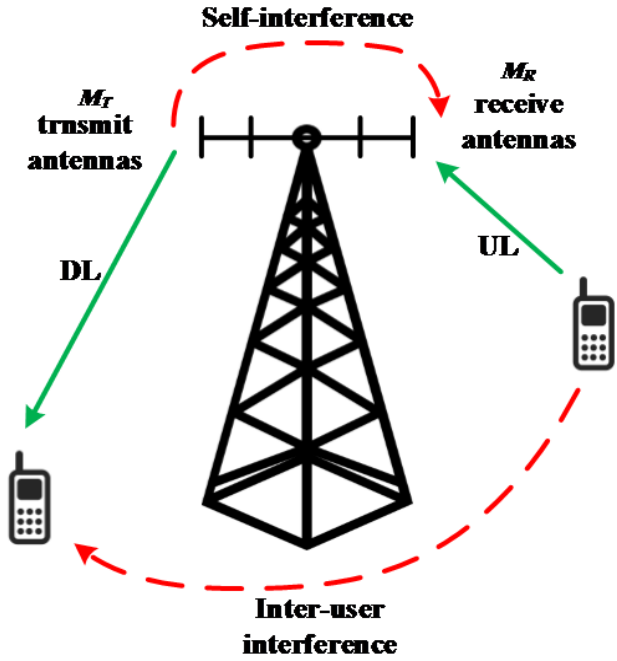}
	\caption{FD multi-antennas base station}
	\label{fig:1}
\end{figure}

The system model according to Fig. \ref{fig:1} is comprised of an FD base station with $M_T$ transmit and $M_R$ receive antennas. In general, $K_D$ user terminals on the DL path receive data from the base station and $K_U$ terminals on the UL path is sending data to the base station. Terminals work as HD. That is, each terminal sends its data to the base station at a certain time slot and in another time slot, it receives its data from the base station. The time slot for sending and receiving for each user are not the same. Therefore, the FD base station can serve $K_D$ user terminals on the DL path and $K_U$ user terminals on the UL path, at the same time.
The user`s multiple access will be considered as one of the most common and orthogonal methods such as orthogonal frequency division multiple access (OFDMA). Since the function of the base station is FD, so the base station provides each of its time-frequency resource blocks to one DL and one UL user simultaneously. For simplicity and without losing generality, we assume that the base station has allocated its first time-frequency resource block to the first user terminal on the DL path and the first user terminal on the UL path, and other time-frequency blocks are available for other users. Because multiple access of the network is considered to be orthogonal, users of different blocks do not interfere with each other. For this reason, in Fig. \ref{fig:1}, only one user in the DL path and one user in the UL path is considered. From hereafter, we only consider the first time-frequency resource block and the first DL and UL user terminals. Because the base station works as FD, the self-interference enters from the transmitter to the receiver. Also, due to the fact that the UL user transmits in the same time-frequency resource block which the DL user receives its own data, there is an intra-user interference between these two terminals. In Fig. \ref{fig:1}, self-interference and intra-user interference are specified as dashed lines. Because the DL and UL users are often away from each other and as the base station are not at high altitudes, the intra-user interference is not as severe as self-interference. Therefore, for simplicity, this interference is not considered in the following.

The base station has $M_{T}$ transmit antennas and the power gains of these channels in the DL path will be shown by $\{h_1,h_2,...,h_{M_T}\}$ . Similarly, the channel power gains between the UL user and $M_R$ receive antennas is $\{g_1,g_2,...,g_{M_R}\}$. In the interference channel, there are $M_R \times M_T$ paths and their power gains can be presented as
\[
\mathbf{A}=
\begin{bmatrix}
\alpha_{11}       & \alpha_{12} & \dots  & \alpha_{1M_T} \\
\alpha_{21}       & \alpha_{22} & \dots & \alpha_{2M_T} \\
\vdots & \vdots& \ddots&\vdots \\
\alpha_{M_R1}       & \alpha_{M_R2} &  \dots & \alpha_{M_RM_T}
\end{bmatrix}
\]
The spatial channels of the interference and the DL and UL paths are uncorrelated and Rayleigh with zero mean and unit variance \cite{Ref16}. Therefore, the power gains of these channels have an exponential distribution \cite{Ref22}. The intra-user interference has been neglected too. In each use of the channel, the base station selects one antenna at the transmitter and one antenna at the receiver. After antenna selection, the channel power gain at the DL and UL will be shown with $h$ and $g$ respectively and the power gain of the interference channel is $\alpha$ . If the received power by the user terminal in the DL path is equal to $P_D$ and the received power by the base station in the UL path is equal to $P_U$, the signal power to noise plus interference power ratio (SINR) in the DL and UL path are respectively \cite{Ref18}
\begin{equation}\label{eq:1}
\gamma_D =\frac{P_Dh}{\sigma_0^2}
\end{equation}
\begin{equation}\label{eq:2}
\gamma_U =\frac{P_Ug}{\sigma_0^2+\eta{P_D}\alpha}
\end{equation}
where $\sigma_0^2$ is the noise power and $0 \leq{\eta}\leq1$ indicates the self-interference cancellation factor in the FD receiver. Note that $\eta=0$ shows the full elimination of the interference and $\eta=1$ means complete leakage of the interference within the FD receiver. After selecting the suitable antenna in the DL and UL paths, and according to \eqref{eq:1} and \eqref{eq:2}, the sum throughput rate of the DL and UL users can be written as \cite{Ref18} 
\begin{equation}\label{eq:3}
C_T =C_D(1-p_{o,D})+C_U(1-p_{o,U}).
\end{equation}
Here, $p_{o,D}$ and $p_{o,U}$ represent outage probability in the DL and UL paths respectively
\begin{equation}\label{eq:4}
p_{o,D}=\mathbb{P}(\gamma_D<\gamma_{D,T})
\end{equation}
\begin{equation}\label{eq:5}
p_{o,U}=\mathbb{P}(\gamma_U<\gamma_{U,T})
\end{equation}
and $\mathbb{P}(.)$ indicates the probability of an event. Also, $C_D$ and $C_U$ show the outage capacity in the DL and UL and $\gamma_{D,T}$ and $\gamma_{U,T}$ are threshold level on the SINR in the DL and UL paths. It is clear that the outage capacity can be written as
\begin{equation}\label{eq:6}
C_D=\log_2(1+\gamma_{D,T})
\end{equation}
\begin{equation}\label{eq:7}
C_U=\log_2(1+\gamma_{U,T})
\end{equation}
In \cite{Ref18}, two single-objective criteria for antenna selection are considered. One of them selects one transmit and one receive antenna which maximizes the DL and UL gains and the other selection criterion minimize the interference channel gain by selecting one antenna at the transmitter and one antenna at the receiver. These two criteria are explained more in the following.

\subsection{Maximum Gain Criterion in DL and UL Path}
In this criterion, one transmit antenna at the base station with the maximum DL channel gain is selected. In this way  
\begin{equation}\label{eq:8}
h=\max\{h_1,h_2,...,h_{M_T}\}
\end{equation}
Similarly, one receive antenna with the maximum UL channel gain is selected. Thus,
\begin{equation}\label{eq:9}
g=\max\{g_1,g_2,...,g_{M_R}\}
\end{equation}
When the transmit and receive antennas are determined according to \eqref{eq:8} and \eqref{eq:9}, the interference channel gain $\alpha$ will also be determined. Similar to \cite{Ref18}, this criterion is abbreviated as MM-AS (max-max antenna selection) in the following.

\subsection{Minimum Gain Criterion on Interference Path}
In this criterion, one transmit and one receive antenna are selected to minimize the self-interference channel gain. So that, the minimum interference will be entered to the receiver. Therefore, 
\begin{equation}\label{eq:10}
\alpha=\min\{\alpha_{11},\alpha_{21},...,\alpha_{M_R1},\alpha_{12},\alpha_{22},...,\alpha_{M_R2},...\}
\end{equation}
In other words, $\alpha$ is equal to the smallest element of the matrix $\mathbf{A}$. If the element on row $i$ and column $j$ of the matrix $\mathbf{A}$ have the minimum value, it means that the $i$-th receive antenna ($i=1,2,...,M_R$) and the $j$-th transmit antenna ($j=1,2,...,M_T$) at the base station will be used in the UL and DL paths. When transmit and receive antennas are determined, the DL and UL channel gains, $h$ and $g$ are also specified. In the following, this criterion is abbreviated as LI-AS (loop-back interference antenna selection) \cite{Ref18}.

\section{Multi-objective Criterion for Antenna Selection}\label{sec:multi_objective_criterion}
In \cite{Ref18}, it has been shown by analysis and simulation that when the DL throughput rate is more important, MM-AS is a more appropriate criterion compared to the LI-AS. In contrast, when the UL throughput rate is more important, LI-AS is more appropriate than the MM-AS. This behavior is also justifiable. In the MM-AS method, self-interference is not considered at the base station. Therefore, the MM-AS criterion is suitable for a DL user who is not involved with self-interference. Similarly, in the LI-AS method, the antenna selection is just based on the self-interference of the base station. Therefore, this criterion is suitable for the UL users. Note that, the UL user`s signal is received by interference due to the FD function of the base station.

From the viewpoint of the network and the base station, maximizing the sum throughput rate of both DL and UL users is more important. Therefore, the best antenna for selecting in the DL and UL paths are those antennas that maximize the DL and UL gains and minimize the interference gain as well. In other words, in this situation we face a multi-objective problem. So the transmit and receive antennas which are selected must provide the three goals. These three goals are the maximization of the DL gain, the maximization of the UL gain and the minimization of the interference gain. Therefore, the problem of antenna selection can be described as a multi-objective optimization problem according to \eqref{eq:11}. In \eqref{eq:11}, $i$ and $j$ represent the antenna index at the receiver and transmitter of the base station and $h$ demonstrates the power gain of the channel in the DL path, $g$ presents the power gain of the channel on the UL path and $\alpha$ shows the power gain of the interference path. As it is clear, \eqref{eq:11} has taken all three optimization \eqref{eq:8}, \eqref{eq:9} and \eqref{eq:10} into consideration. Hence, this problem is called multi-objective \cite{Ref23,Ref24}.
\begin{equation}\label{eq:11}
\begin{cases}
\displaystyle\max_{i,j}~h&\\
\displaystyle\max_{i,j}~g&\\
\displaystyle\min_{i,j}~\alpha&
\end{cases}
\end{equation}
The multi-objective optimization problem often does not have a unique answer. This means that there is no definite choice of transmit and receive antennas that meets all three objectives in \eqref{eq:11} at the same time. Therefore, in multi-objective problems, we have a set of solutions instead of a unique answer. This set is called the optimal Pareto points (see \cite{Ref23,Ref24} for a more precise description and mathematical definition of the optimal Pareto points). In multi-objective optimization problems, one of the optimal Pareto responses is usually used as the final solution of the problem.

Multi-objective optimization problem solving methods are very diverse. Among these methods, the weighted sum method is simple and of course very efficient \cite{Ref23}. In this method, the objective functions of the problem are combined with different weights and create a new objective function. In this way, the multi-objective problem changes to a common single-objective problem, which is simple to solve \cite{Ref23}. According to these explanations, we can transform the multi-objective optimization problem \eqref{eq:11} into a single-objective optimization problem in \eqref{eq:12} which  $0\leq w_1 \leq 1$  , $0\leq w_2 \leq 1$ and $0\leq w_3 \leq 1$ indicate weights in the weighted sum method. These weights determine the importance of each of the objective functions of the optimization problem \eqref{eq:11}.
\begin{equation} \label{eq:12}
\displaystyle\min_{i,j}~-w_1h-w_2g+w_3\alpha.
\end{equation}
In \eqref{eq:11}, since the first two targets were considered to be maximal $h$ and $g$, they were entered in \eqref{eq:12} with a negative sign. Often the importance of the DL and UL users are the same. Therefore, we can assume that the coefficients of $w_1$ and $w_2$ are equal. Also, because the total weight is considered to be one, we can rewrite \eqref{eq:12} in a simpler way and in the form of 
\begin{equation} \label{eq:13}
\displaystyle\min_{i,j}~-\left(\frac{1-w}{2}\right)h-\left(\frac{1-w}{2}\right)g+w\alpha.
\end{equation}
According to \eqref{eq:13}, one antenna will be selected at the transmitter and receiver of the base station to minimize \eqref{eq:13}. As it was said, weights determine the importance of each of the objective functions and this importance may change in good and bad channel conditions. In the problem \eqref{eq:13}, when $w$ becomes larger, the importance of the interference channel and interference cancellation become more important. As the same way, the performance of the system becomes more similar to the LI-AS. Eventually when $w=1$, the optimization problem \eqref{eq:13} is exactly equivalent to the LI-AS criterion. In contrast, when $w$ decreases, the importance of the interference channel becomes less and less, and the importance of the direct channel increases in the DL and UL paths. In this case, the performance of the system with multi-objective antenna selection is similar to the MM-AS and in a certain degree of $w=0$ , exactly equals the MM-AS criterion. The precise adjustment of the parameter $w$ in solving the multi-objective optimization problem does not follow a definite method, and is often carried out by an expert and according to the conditions of the problem \cite{Ref23,Ref24}. In this paper, in order to simplify the multi-objective problem solving, experimentally and in terms of trial and error, a mathematical relation is proposed for setting $w$ .This experimental relation is confirmed in the next section by simulations.

\section{Performance Evaluation and Simulation Results}\label{sec:simulation_results}
In this section, the performance of an FD base station with a new antenna selection criterion is reviewed and Monte Carlo simulation results are presented. For this purpose, we have
$$\frac{P_D}{\sigma_0^2}=\frac{P_U}{\sigma_0^2}=\gamma_0$$
and $\gamma_0$ shows the average signal power to noise power ratio (SNR) in the DL and UL paths. Also $\gamma_{D,T}$=$\gamma_{U,T}$=10dB is assumed. For Monte Carlo simulation, 100,000 samples have been used in each simulation.
\begin{figure}
	\includegraphics[width=1.0\textwidth]{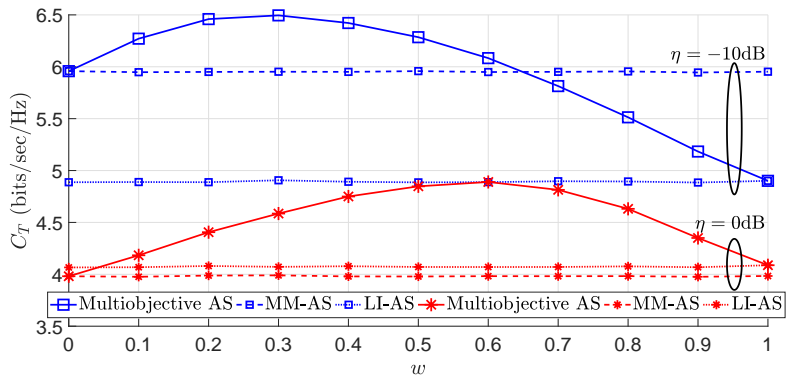}
	\caption{Comparison of the new multi-objective criterion for antenna selection with the two previous criteria (MM-AS and LI-AS) for relatively low and relatively high interference}
	\label{fig:2}
\end{figure}
First, in Fig. \ref{fig:2}, the sum throughput rate for the three different criteria (MM-AS, LI-AS and Multi-objective-AS) is presented versus the weight $w$ . In this figure, $M_T=4$ , $M_R=4$ and $\gamma_0$=15dB. In the blue curves with a square marker, $\eta=-10$dB is assumed and in the red curves which are characterized by a stellar marker, $\eta=0$dB . In Fig. \ref{fig:2}, the sum throughput rate with MM-AS and LI-AS can be compared with each other. For relatively low interference ($\eta=-10$dB), the throughput rate with MM-AS is higher than the throughput rate with LI-AS. But in the relatively high interference ($\eta=0$dB), the throughput rate with the LI-AS is higher than the MM-AS. Therefore, in varying degrees of interference, the MM-AS and LI-AS criteria have different performance. The reason for this difference, as already explained, is that, in the MM-AS criterion for selecting an antenna, no attention is paid to interference channel and the antennas that have the maximum gain on the DL and UL paths are selected. This criterion is suitable for a low interference situation. But when the interference is intense, the selection of the antenna with the LI-AS criterion performs better because it pays attention to the interference channel and selects antennas that have the least gain in the interference channel. The multi-objective antenna selection criterion, discussed in this article, tries to combine both criteria and improves performance in all situations. In Fig. \ref{fig:2}, it is clear that the multi-objective antenna selection with $w=0$ and $w=1$ performs similar to those of MM-AS and LI-AS. However, the most important point in Fig. \ref{fig:2} is the performance improvement of the criterion presented in this paper, in comparison with the MM-AS and LI-AS. When the interference is relatively low ($\eta=-10$dB), the multi-objective antenna selection has a better sum throughput rate than either MM-AS or LI-AS for $0\leq w \leq0.65$ . When $w\approx0.3$, this sum throughput rate will be maximized. When the interference is relatively intense ($\eta=0$dB), the multi-objective antenna selection for $0.04\leq w \leq1$ has a better throughput than the previous two criteria and when $w\approx0.6$, the throughput rate will be maximized. Therefore, by setting the $w$ parameter appropriately, selecting the antenna with a new multi-objective criterion will certainly have a better performance in comparison with the common antenna selection methods. The parameter $w$ specifies the importance of the different objective functions in weighted sum method and the appropriate value of this parameter depends on the different conditions of the channel.
\begin{figure}
	\includegraphics[width=1.0\textwidth]{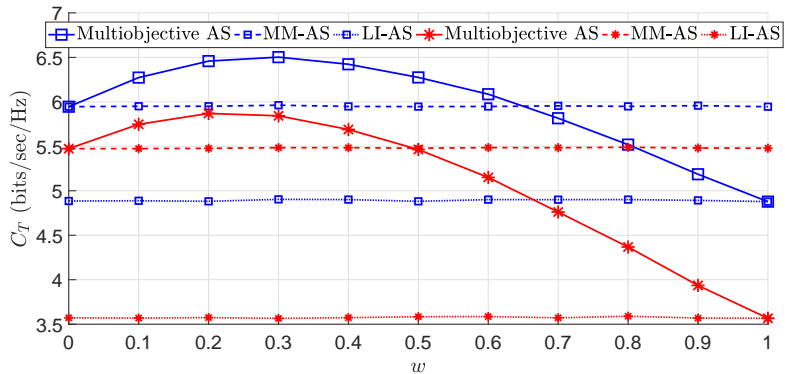}
	\caption{Comparison of the new multi-objective criterion for antenna selection with the two previous criteria (MM-AS and LI-AS) for different values of the average SNR}
	\label{fig:3}
\end{figure}
In Fig. \ref{fig:3} we have plotted the sum throughput rate again for the three different criteria based on $w$. As in the previous figure, $M_T=4$ , $M_R=4$ and here  $\eta=-10$dB is assumed. The figure is drawn for two different values $\gamma_{0}=15$dB (blue curves with square markers) and $\gamma_{0}=12$dB (red curves with stellar markers). As in Fig. \ref{fig:2}, the curves are comparable in different aspects. The most important point is the difference in the performance of the multi-objective criterion when $\gamma_{0}$ is changed. When $\gamma_{0}=15$dB, the sum throughput rate reaches its maximum value for $w\approx0.3$ and when $\gamma_{0}=12$dB, the sum throughput rates for $w\approx0.2$ is maximized. This change is also seen in Fig. \ref{fig:2} for different values of $\eta$. So, the correct setting of $w$ is of great importance, according to the conditions of the channel such as $\gamma_0$ and $\eta$ .

\begin{figure}
	\includegraphics[width=1.0\textwidth]{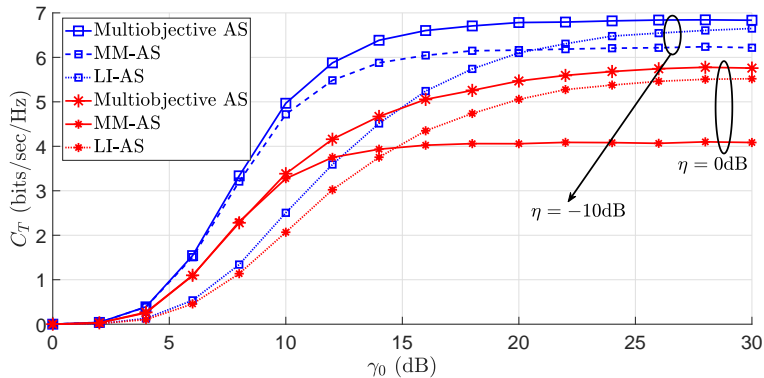}
	\caption{Comparison of the new multi-objective antenna selection criterion with two previous criteria (MM-AS and LI-AS) in different channel conditions when the number of transmit and receive antennas is 4}
	\label{fig:4}
\end{figure}

\begin{figure}
	\includegraphics[width=1.0\textwidth]{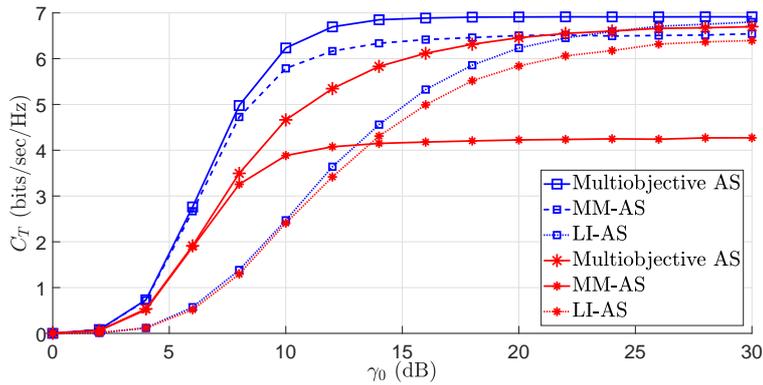}
	\caption{Comparison of the new multi-objective antenna selection criterion with two previous criteria (MM-AS and LI-AS) in different channel conditions when the number of transmit and receive antennas is 8}
	\label{fig:5}
\end{figure}

The proper setting of the parameter $w$ is often carried out empirically by an expert \cite{Ref23,Ref24}. To simplify the proper adjustment of this parameter, with trial and error, a suitable empirical value for $w$ is proposed as 
\begin{equation}
w=0.5\eta^{0.301}+0.02\breve{\gamma_0}-0.3.
\end{equation}
In this case, $\breve{\gamma_0}$ shows the average SNR in dB. To verify the validity of this experimental equation, various simulations have been carried out for different channel and system parameters such as $\eta$, $\gamma_0$, $\gamma_{D,T}$, $\gamma_{U,T}$, $M_T$ and $M_R$. In all of these cases, the multi-objective selection criterion has had a better performance than the two criteria of MM-AS and LI-AS. Two of these simulations are shown in Fig. \ref{fig:4} and \ref{fig:5}, and the rest are not drawn to prevent repetition. Note that equation \eqref{eq:13} is empirically considered for valuing $w$ and has been confirmed by various simulations. Since the weighted sum method is a very simple technique for solving the multi-objective optimization problem, having a proper estimate for $w$ can solve the multi-objective problem quite easily and straightforwardly. However, other methods that do not need to estimate the $w$ parameter can also be used to solve the multi-objective optimization problem.
In Fig. \ref{fig:4}, the sum throughput rate is plotted according to $\gamma_0$. In the blue curves with a square marker, $\eta=-10$dB , and in the red curves with a stellar marker, $\eta=0$dB is assumed. As before, the number of transmit and receive antennas is $M_T=4$, $M_R=4$ and one transmit and one receive antenna is selected in the base station in each time the channel is used. In order to calculate the sum throughput rate with the multi-objective antenna selection criterion, the parameter $w$ is established in accordance with \eqref{eq:13}. As it is clear, the function of the new antenna selection in all values of $\gamma_0$ and for both values of $\eta$ is better than the performance of the commonly used MM-AS and LI-AS. For example, for $\eta=-10$dB and in $\gamma_0=10$dB , the sum throughput rates in the proposed method is 2.46bits/sec/Hz better than the throughput rate in the LI-AS method. Also, for $\eta=-10$dB and in $\gamma_0=20$dB , the sum throughput rate of the multi-objective method is 0.62bits/sec/Hz higher than the sum throughput rate in the MM-AS method. The same comparison can be made in severe interference situations. For $\eta=0$dB and in $\gamma_0=10$dB, the sum throughput rate in the proposed method is 1.22bits/sec/Hz better than the sum throughput rate in the LI-AS method. Also, for $\eta=0$dB and $\gamma_0=20$dB, in the multi-objective method, the sum throughput rates is 1.41bits/sec/Hz higher than the sum throughput rate of the MM-AS.

For having another comparison, similar to Fig. \ref{fig:4}, the sum throughput rate is plotted versus $\gamma_0$ for $M_T=8$ and $M_R=8$ in Fig. \ref{fig:5}. Here again, the parameter $w$ is adjusted according to \eqref{eq:13}. The performance of the new antenna selection in all situations is better than the conventional one. Comparing Fig. \ref{fig:4} and \ref{fig:5}, it is clear that the sum throughput rate in Fig. \ref{fig:5} reaches its maximum value much faster than Fig. \ref{fig:4}. For example, for $\gamma_0=10$dB and $\eta=-10$dB , in Fig. \ref{fig:4}, the sum throughput rate is about 4.97bits/sec/Hz and in Fig. \ref{fig:5}, this value is approximately 6.23bits/sec/Hz. The reason for the increase is that in Fig. \ref{fig:5} the number of available antennas for selection is higher.

\begin{figure}
	\includegraphics[width=1.0\textwidth]{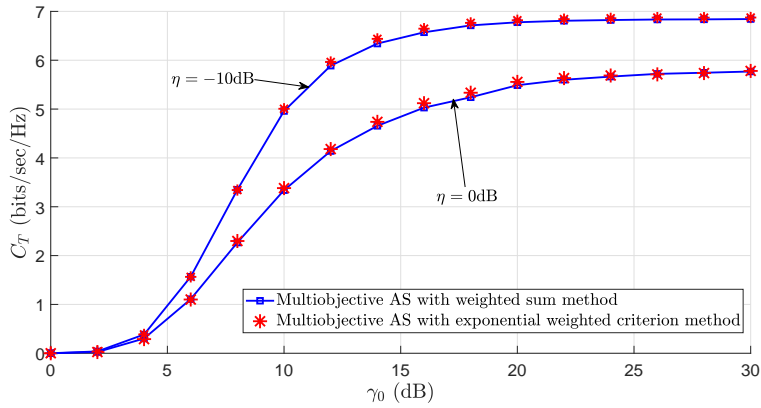}
	\caption{Comparison of weighted sum method and exponential weighted criterion method for multi-objective optimization solution}
	\label{fig:6}
\end{figure}

Finally, in Fig. \ref{fig:6} the sum throughput rate is plotted versus $gamma_0$ for $M_T=4$, $M_R=4$ and $\eta=-10$dB, $\eta=0$dB. Here, we use two different methods to solve the multi-objective optimization and compare the obtained results. First, the solution with the weighted sum method is depicted with blue curves and square markers in the figure. Then, the second solution with the exponential weighted criterion method \cite{Ref23,Ref24} is plotted with red stellar markers. The exponential weighted criterion method has a similar parameter $w$ and this parameter is adjusted to its proper value by trial and error. However, in weighted sum method, $w$ can be set by \eqref{eq:13}. The performance of two solutions are nearly the same. Although the weighted sum method is more simple to use and simulate.
\section{Conclusion}\label{sec:conclusion}
In this paper, a new criterion for selecting an antenna in an FD base station is presented. In this new criterion, antenna selection has been modeled as a multi-objective optimization problem. In this case, multi-objective optimization is considered simultaneously with the maximization of the gain in the DL and UL paths along with the minimization of the gain in the interference channel and is solved by the weighted sum method. An experimental equation is also proposed for parameter setting to simplify using weighted sum method. Simulation results show performance improvement in the new antenna selection method than the previous ones.

\bibliographystyle{unsrt}
\bibliography{References}
\end{document}